\documentclass{article}

\usepackage[preprint]{neurips_2021}

\usepackage[utf8]{inputenc} 
\usepackage[T1]{fontenc}    
\usepackage{hyperref}       
\usepackage{url}            
\usepackage{booktabs}       
\usepackage{amsfonts}       
\usepackage{nicefrac}       
\usepackage{microtype}      
\usepackage{xcolor}         
\usepackage{wrapfig}
\usepackage{bm}
\usepackage{amsmath}
\usepackage{amssymb}
\usepackage{amsthm}
\usepackage{amsmath}
\usepackage[inline]{enumitem}
\usepackage{algorithm}
\usepackage[noend]{algpseudocode}
\usepackage{graphicx}
\usepackage{caption}
\usepackage{subcaption}
\usepackage{diagbox}
\usepackage{multirow}
\usepackage{apptools}
\usepackage{diagbox}
\usepackage{makecell}

\newtheorem{proposition}{Proposition}
\newtheorem{definition}{Definition}

\newtheorem{problem}{Problem}

\AtAppendix{\counterwithin{lemma}{section}}
\AtAppendix{\counterwithin{theorem}{section}}
\AtAppendix{\counterwithin{proposition}{section}}
\AtAppendix{\counterwithin{corollary}{section}}

\definecolor{azure}{rgb}{0.0, 0.5, 1.0}
\definecolor{brandeisblue}{rgb}{0.0, 0.44, 1.0}
\definecolor{darkpastelgreen}{rgb}{0.01, 0.75, 0.24}
\definecolor{darkpastelpurple}{rgb}{0.59, 0.44, 0.84}
\definecolor{darktangerine}{rgb}{1.0, 0.66, 0.07}
\definecolor{debianred}{rgb}{0.84, 0.04, 0.33}
\definecolor{hanpurple}{rgb}{0.32, 0.09, 0.98}
\definecolor{deepcerise}{rgb}{0.85, 0.2, 0.53}
\definecolor{emerald}{rgb}{0.31, 0.78, 0.47}
\definecolor{fuchsia}{rgb}{1.0, 0.0, 1.0}
\definecolor{flamingopink}{rgb}{0.99, 0.56, 0.67}
\definecolor{lightseagreen}{rgb}{0.13, 0.7, 0.67}
\definecolor{mayablue}{rgb}{0.45, 0.76, 0.98}

\newcommand{\PP}{\mathbb{P}}
\newcommand{\EE}{\mathbb{E}}

\newcommand{\X}{\mathcal{X}}
\newcommand{\FF}{\mathcal{F}}

\newcommand{\R}{\mathbb{R}}

\newcommand{\T}{\mathcal{T}}

\title{Calibration of Derivative Pricing Models: a Multi-Agent Reinforcement Learning Perspective}

\author{
Nelson Vadori\thanks{correspondence to: nelson.n.vadori@jpmorgan.com}
 \\
J.P. Morgan AI Research\\
}

\begin{document}

\maketitle

\begin{abstract}
One of the most fundamental questions in quantitative finance is the existence of continuous-time diffusion models that fit market prices of a given set of options. Traditionally, one employs a mix of intuition, theoretical and empirical analysis to find models that achieve exact or approximate fits. Our contribution is to show how a suitable game theoretical formulation of this problem can help solve this question by leveraging existing developments in modern deep multi-agent reinforcement learning to search in the space of stochastic processes. Our experiments show that we are able to learn local volatility, as well as path-dependence required in the volatility process to minimize the price of a Bermudan option. Our algorithm can be seen as a particle method \textit{\`{a} la} Guyon \textit{et} Henry-Labordere where particles, instead of being designed to ensure $\sigma_{loc}(t,S_t)^2 = \EE[\sigma_t^2|S_t]$, are learning RL-driven agents cooperating towards more general calibration targets.
\end{abstract}

\section{Introduction}
\label{sec1}

Throughout, we consider a single equity asset which price at time $t$ we denote $S_t$, but our methods can be extended straightforwardly to the multi-dimensional case. Let $S:=(S_t)_{t \in [0,T]}$ be the (stochastic) price path process over some finite time horizon $[0,T]$ on a complete probability space $(\Omega, \FF, \PP)$. We denote $\EE$ the expectation with respect to $\PP$. An option is simply a real-valued random variable $\X$ which can, in general, depend on $\mbox{Law}(S)$, the distribution of the path $S$. Among the most liquid options are call options (also referred to as "vanillas"), for which $\X:=(S_{t} - k)^+:= \max(S_{t} - k, 0)$, where the option expiry $t$ and strike $k$ are fixed. We consider continuous-time diffusion models of the form $dS_t = S_t \sigma_t dW_t$, where $W$ is a Brownian motion, and we let the stochastic volatility process $\sigma$ be very general for now, for example it could be path-dependent, function of extra sources of randomness, or rough \cite{guyonpv, roughvol}. Such models are uniquely determined by $\sigma$, so we will call $\sigma$ a model, and denote $\EE \X$ the price of option $\X$ under model $\sigma$. As it is usual to do so, we consider, without loss of generality, zero interest rate and dividends, so that $S$ coincides with the forward price process. 

One of the most fundamental questions arising in quantitative finance is the following: 
\begin{problem}
\label{p1}
\textit{Given a set of options $\mathcal{O}$ with market prices $\mathcal{P}_{mkt}$, find a continuous-time diffusion $\sigma$ such that model prices $\mathcal{P}_{\sigma}$ of $\mathcal{O}$ are equal to $\mathcal{P}_{mkt}$}.
\end{problem}  

In the case where $\mathcal{O}$ consists of a single call option, the well-known Black \& Scholes model is a solution to problem \ref{p1}. When $\mathcal{O}$ consists of many call options, the Dupire formula and associated local volatility model is a solution. Actually, it is now well-known that one can take any $\sigma$ and localize it using the Gy\"{o}ngy Markovian projection formula $\sigma_{loc}(t,S_t)^2 = \EE[\sigma_t^2|S_t]$, resulting in the localized volatility $\EE[\sigma_t^2|S_t]^{-\frac{1}{2}} \sigma_{loc}(t,S_t) \sigma_t$ and the associated class of stochastic local volatility models (SLV) \cite{guyonpa}. For more general $\mathcal{O}$, though, solving problem \ref{p1} is hard.

\begin{figure}[ht]
    \centering
    \begin{subfigure}[b]{0.49\textwidth}
        \centering
        \includegraphics[width=\textwidth]{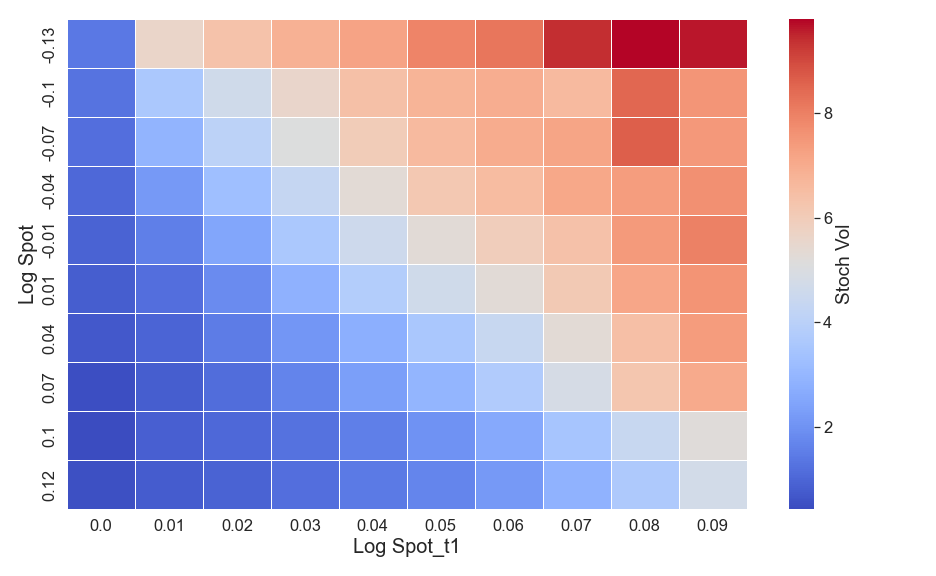}
    \end{subfigure}
    \hfill
    \begin{subfigure}[b]{0.49\textwidth}
        \centering
        \includegraphics[width=\textwidth]{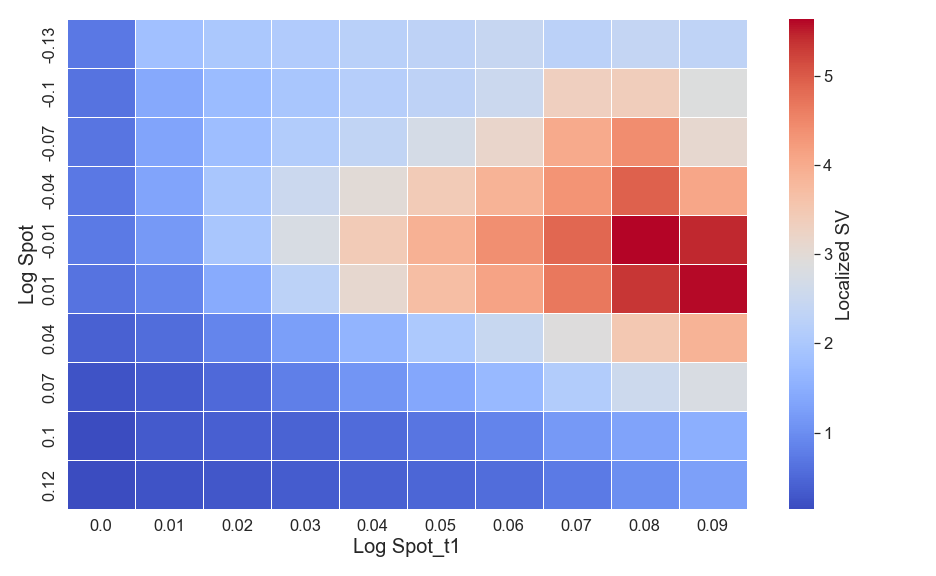}
    \end{subfigure}
    \caption{Bermudan option with exercise dates $t_1=21$ days, $t_2=51$ days. (Left) average volatility $\sigma_t$ and (right) average localized volatility $\EE[\sigma_t^2|S_t]^{-\frac{1}{2}} \sigma_t$ as a function of $\ln S_{t_1}$ and $\ln S_t |_{t>t_1}$ (averaged over $t>t_1$). Strike $\ln k_1=0$. As intuited, we learn a volatility 
    increasing in $S_{t_1}$, decreasing in $S_{t}|_{t>t_1}$. SPX, August 1st 2018.}
    \label{figb2}
\end{figure}

As an example, let us consider the case where $\mathcal{O}$ consists of two types of options: european call options for all strikes (smile) and all dates $t \in [t_1,t_2]$, and a Bermudan option with exercise dates $t_1$, $t_2$, for which we have $\X:= \max \{(S_{t_1} - k_1)^+,\EE_{t_1}[(S_{t_2} - k_2)^+]\}$, where $k_1$, $k_2$ are fixed and $\EE_{t_1}$ denotes the expectation conditional on $\FF_{t_1}$, the information available up to $t_1$. Due to $\EE_{t_1}$ and the "$\max$" non-linearity, the Bermudan option price depends on the forward smile $t_1 \to t_2$ and cannot be replicated with vanillas. Such option price is lower bounded by the maximum of the two european call options $(t_1,k_1)$, $(t_2,k_2)$, which we call maximum european: $\EE \X \geq \max \{\EE (S_{t_1} - k_1)^+,\EE(S_{t_2} - k_2)^+\}$ \footnote{remember that $\EE [\EE_{t_1} [\cdot]] = \EE[\cdot]$.}. Assume that $\mathcal{P}_{mkt}$ is close to its lower bound, so that our goal is to find $\sigma$ so as to minimize $\mathcal{P}_{\sigma}$. First, notice that $\mathcal{P}_{\sigma}$ equals the maximum european means that if the maximum european is on $t_1$, we should never exercise at $t_2$ for any path $\omega$, and vice-versa if the maximum european is on $t_2$. One can make the following heuristic reasoning to get the intuition that $\sigma_t$ for $t \in [t_1,t_2]$ must remember the value of $S_{t_1}$, and hence be path-dependent, if we hope to minimize $\mathcal{P}_{\sigma}$. If the maximum european is on $t_2$, you need on paths $\omega$ for which $S_{t_1}(\omega)$ is high, a high volatility $\sigma_t(\omega)$ on $[t_1,t_2]$ so that you do not exercise at $t_1$. If the maximum european is on $t_1$, the conclusion is the same, although the reasoning is different: you need to allocate high volatility on paths where $S_{t_1}(\omega)$ is high, since doing so on paths where it is low will clearly result in not exercising at $t_1$. Note that we use the terminology "allocate volatility", since the (unconditional) aggregated variances $\EE \ln^2 S_{t_1}$ and $\EE \ln^2 S_{t_2}$ are fixed by vanilla market smiles. Our informal reasoning will be validated by our learnt $\sigma$ in figures \ref{figb2}-\ref{figb3}.

It has been shown in \cite{guyonpv} that volatility can, in theory, be path-dependent while preserving perfect fit to vanilla smiles. Even then, the difficulty lies in choosing the specific form of path dependence. Guyon provides examples of how to choose it in a heuristic way so as to achieve specific features, for example on the forward smile. The now popular class of rough volatility models \cite{roughvol} also falls into the path-dependent category. The novel work of \cite{deepc1,deepc2} let $\sigma$ be a neural network and calibration is performed via deep learning. Among other things, this opens the door to learning complex path-dependence of the volatility process, even though this aspect is not explored in these works. In particular, this approach requires storing the whole computational graph, which, due to memory considerations, limits the number of time steps and Monte Carlo trajectories that one can use to simulate the diffusion process. For this reason, both these work consider a small number of maturities to be calibrated (4-6), one neural network per maturity, and train one maturity at a time, since \textit{"training all neural networks at every maturity all at once, makes every step of the gradient algorithm used for training computationally heavy"}. Note that the same limitation has been identified in the \textit{learning to optimize} literature, where the goal is to replace the traditional optimization gradient step by a neural network \cite{bl2o,bl2o2}. There, because of the aforementioned memory bottleneck, one typically trains on short time horizons, or optimization trajectories.

\textbf{Contribution.} We introduce a novel game theoretical formulation of the calibration problem \ref{p1} where individual particles acting locally cooperate towards the calibration objective at aggregate level. We introduce a numerical method to solve it by leveraging existing developments in modern deep multi-agent reinforcement learning to search in the space of stochastic processes. The latter is a reinforcement learning driven alternative to the aforementioned deep learning based approaches, which in particular doesn't require storing the computational graph and as such doesn't suffer from memory bottlenecks. This allows us to "train" all maturities at once using the concept of rewards obtained along agent trajectories, which is required in particular to handle the Bermudan option case discussed above, or to calibrate to a continuum of vanilla smiles in one shot, as we do in our experiments. In essence, our idea is to see Monte Carlo trajectories as players participating in a cooperative game, where player $i$'s action at time $t$ is the choice of the volatility $\sigma^i_{t}$ to be used on trajectory $i$ over $[t,t+dt]$, and where the common reward that players try to minimize is the calibration error, which is, in game theoretical terms, a function of \textit{all} players' individual actions. Our experiments show that we are able to learn local volatility, as well as the path-dependence required to minimize the price of the Bermudan option previously discussed, matching our initial intuition. Finally, on the multi-agent reinforcement learning side, we introduce a 
new technique to handle the case of a very large number of players ($\sim 10^5-10^7$), that turned out to be crucial in making our method work, where \textit{policy exploration} across all players is obtained via interpolation of a few \textit{basis players} ($\sim 10^2$).

\section{Game theoretical formulation of the calibration problem}
\label{sec2}

As stated in introduction, we let $S:=(S_t)_{t \in [0,T]}$ be the stochastic price path process over some finite time horizon $[0,T]$ on a complete probability space $(\Omega, \FF, \PP)$ and consider continuous-time diffusion models of the form:
\begin{align}
\label{eqspot}
    dS_t = S_t \sigma_t dW_t,
\end{align}
where $W$ is a Brownian motion and $\sigma$ is allowed to be a very general stochastic process. 

We start by giving a concise exposition of the key concepts in Markov decision processes (single-agent case) and stochastic games (multi-agent case), that we will use to develop our approach. Reinforcement learning (RL) and its multi-agent counterpart (MARL) are used as numerical methods to solve the latter, which we will focus on in section \ref{sec3}. We refer to \cite{Sutton} for a presentation of the core concepts in RL, and \cite{jaim,rxu} illustrate some of its applications to finance.

Under a Markov decision process framework, an agent evolves over a series of discrete time steps $t \in [0,T]$. At each $t$, it observes some information called its "state" $x_t$, takes an action $a_t$, and as a consequence, transitions to a new state $x_{t+1}$ as well as gets a reward $r(x_t,a_t,x_{t+1})$, where the transition dynamics $\T(x_{t+1}|x_t,a_t)$ is called the "environment" and denotes the probability to reach $x_{t+1}$ conditional on $x_t,a_t$. The design of the state $x_t$ is quite flexible, in particular one can include time $t$ as well as path-dependent information. This is even more true since the recent developments around neural networks which can be fed a large amount of information, and aforementioned path-dependent data can be encoded by means of the hidden state of a LSTM (long short-term memory). The agent samples its actions according to the stochastic policy $\pi(a_t|x_t)$ representing the probability density function of $a_t$ conditionally on $x_t$, and its goal is simply to maximize its expected cumulative reward:
\begin{align}
\label{vf}
    \sup_\pi v(\pi) := \EE_\pi \sum_{t=0}^T r(x_t,a_t,x_{t+1}),
\end{align}

where $\EE_\pi$ indicates that $\mbox{Law}(a_t|x_t)=\pi$, and $v(\pi)$ is usually called the value function. We focus here on the finite time horizon case $T<\infty$ and omit the classical discounting of rewards by $\gamma^t$ since we do not need it in the present work. Due to the finite time horizon, we assume that $x_t$ includes time $t$, since it is well-known that solutions of (\ref{vf}) must be time-dependent. A stochastic game is simply the extension of a Markov decision process to $n>1$ players: each agent $i \in [1,n]$ is now allowed to have its own policy $\pi_i \equiv \pi_i(a^i_t|x^i_t)$ and reward function $r_i$, but the latter now depends, in the most general case, on all agents' actions and states $r_i \equiv r_i(\bm{x_t}, \bm{a_t}, \bm{x_{t+1}})$, where we denote vectors in bold $\bm{x_t}:=(x^i_t)_{i \in [1,n]}$, $\bm{a_t}:=(a^i_t)_{i \in [1,n]}$. We consider the fully observable case where each player observes information $\bm{x_t}$ and actions $\bm{a_t}$ from all players at each time $t$. We focus our attention on symmetric cooperative games where:
\begin{itemize}
    \item \textbf{(cooperation)} all agents rewards are the same $r_i \equiv r$.
    \item \textbf{(symmetry)} agents are interchangeable: $r_i(\bm{x_t}, \bm{a_t}, \bm{x_{t+1}})=r_{\rho(i)}(\bm{x_t}^{\rho}, \bm{a_t}^{\rho}, \bm{x_{t+1}}^{\rho})$ for all permutations $\rho$, where the "superscript $\rho$" denotes the permuted vector. In simple terms, this condition means that what matters is "what" is done and "where", but not "who" did it. In mathematical terms, the "what", "where" and "who" are respectively $a^i_t$, $x^i_t$, and $i$.
\end{itemize}

One should think about such games as a set of interacting particles, or ants, acting locally towards a common goal at the aggregate level.

Solving cooperative games in the context of MARL has for example been studied in \cite{Gupta2017-it, nipsFoerster}. Similar to the single-agent case (\ref{vf}), we denote the value function of the symmetric cooperative game:
\begin{align}
\label{vf2}
    v(\bm{\pi}) := \EE_{a^i \sim \pi_i: i \in [1,n]} \sum_{t=0}^T r(\bm{x_t}, \bm{a_t}, \bm{x_{t+1}}).
\end{align}

\begin{definition}
\textbf{(Nash equilibrium)} The vector of players' policies $\bm{\pi^*}:=(\pi^*_i)_{i \in [1,n]}$ of the symmetric cooperative game associated to value function $v$ in (\ref{vf2}) is a Nash equilibrium if:
\begin{align}
v(\pi_i; \bm{\pi^*}_{-i}) \leq v(\bm{\pi^*}) \hspace{3mm} \forall \pi_i, \hspace{1mm}\forall i \in [1,n],
\end{align}
where $v(\pi_i; \bm{\pi^*}_{-i})$ denotes the value of the game associated to $\bm{\pi^*}$, but with $\pi^*_i$ replaced by $\pi_i$.
\end{definition}

In simple terms, no player has an incentive to deviate unilaterally in a Nash equilibrium. 

Back to our calibration problem. Let each Monte Carlo trajectory be a player, so that $n$ is the number of Monte Carlo trajectories, typically very large ($10^5-10^7$). Let player $i$'s action $a^i_t:=\sigma^i_t$, the volatility to use for trajectory $i$ over the timestep $[t,t+1]$. We can more generally allow $\sigma^i_t \sim \varphi(\cdot; a^i_t, x^i_t)$ for some distribution $\varphi$, but we discuss this later. The discretization of the dynamics (\ref{eqspot}) yields:
\begin{align}
\label{lns}
\ln S^i_{t+1} = \ln S^i_{t} -\frac{1}{2} \sigma^{i,2}_t \delta + \sqrt{\delta} \sigma^{i}_t Z^i_t,
\end{align}

where $(Z^i_t)_{\substack{t \in [0,T], i \in [1,n]}}$ is a collection of i.i.d. standard normal random variables and $\delta$ is the time step value, for example $\delta = \frac{1}{252}$ for a time step of one day. The state $x^i_t$ typically contains spot and volatility for trajectory $i$ over a past historical window of size $h$, $(S^i_u, \sigma^i_{u-1})_{u \in [t-h,t]}$, as well as time $t$. It could also contain information relative to other trajectories, to include the McKean-Vlasov case where the diffusion coefficients depend on $\mbox{Law}(S_t)$. As previously mentioned, the now standard technique to encode history of observations, required for example in imperfect information games, is via the hidden state of a LSTM \cite{Gupta2017-it}, but other approaches are possible, for example path signatures \cite{sig}.

We assume that we have $n_t^o$ option cashflows $\{\X^m_t: m \in [1,n_t^o], t \in [0,T]\}$ associated to market target prices $c_t^m \in \R$. Option cashflows are random variables, and each $\X^m_t$ is assumed to be a function of $(\bm{x_t}, \bm{a_t}, \bm{x_{t+1}})$, remembering from our previous discussion that states $x^i_t$ can potentially encapsulate a large amount of information, including path-dependent one. We then define the game reward obtained at $t+1$ as the calibration loss:
\begin{align}
\label{rcalib}
\begin{split}
   & r(\bm{x_t}, \bm{a_t}, \bm{x_{t+1}}) := - \sum_{m=1}^{n_t^o} \omega^m_t \left(\phi^m_t(\widehat{\X}^{m}_{t+1}) - c^m_t\right)^2,\\
   & \widehat{\X}^{m}_{t+1} :=\frac{1}{n}\sum_{j=1}^n \X^{j,m}_{t+1},
\end{split}
\end{align} 
where $\omega^m_t$ are weights chosen by the user (for example related to the Vega of the option), $\widehat{\X}^{m}_{t+1}$ is the Monte Carlo price and $\phi^m_t$ a suitable transformation, for instance the Black \& Scholes implied volatility mapping in the case of a call option, in which case $c^m_t$ would be the market implied volatility. In the case of a call option, we have $\X^m_t:=(S_t-k_m)^+$. In the case of the Bermudan$(t_1,t_2)$, we have $\X_t=0$ for $t<t_2$ and $\X_{t_2}:= \max \{(S_{t_1} - k_1)^+,\EE_{t_1}[(S_{t_2} - k_2)^+]\}$, where the conditional expectation is calculated empirically using all trajectories over $[t_1,t_2]$ using, for instance, American Monte Carlo regression methods \cite{ls}. 

The deep MARL methods we employ in section \ref{sec3} allow to have sparse reward signals as it is the case for the Bermudan option, the key being to use a neural network to approximate the value function, i.e. the expected sum of future rewards, given a particular state $x^i_t$. Such approximation is called a "critic" in RL jargon, and was one of the keys to the success of AlphaGo, Deepmind's pioneering work on the game of Go \cite{SilverHuangEtAl16nature}. 

Even though the true rewards can be sparse like in the Bermudan example, it is actually possible in many cases to construct per-timestep approximations of true rewards. Reward design, or "shaping" is an active area of research in RL. It consists in suitably designing the reward function $r(\cdot)$ to improve learning. In our case, it is perfectly reasonable to define "fake" cashflows $\widetilde{\X}_{t}$ and rewards $\widetilde{r}$ such that cumulative rewards agree:
\begin{align}
\label{rfake}
    \sum_{t=0}^T r(\bm{x_t}, \bm{a_t}, \bm{x_{t+1}}) = \sum_{t=0}^T \widetilde{r}(\bm{x_t}, \bm{a_t}, \bm{x_{t+1}}).
\end{align}
In the Bermudan$(t_1,t_2)$ example, a reasonable choice would be $\widetilde{\X}_{t}:= \max \{(S_{t_1} - k_1)^+,\EE_{t_1}[(S_{t} - \widetilde{k}_{2,t})^+]\}$, where $\widetilde{k}_{2,t}$ can be expressed either in percentage of the forward at time $t$, or in delta space $\widetilde{k}_{2,t}=S_0 \left(\frac{k_2}{S_0}\right)^{\sqrt{\frac{t}{t_2}}}$. We then define the fake rewards as:
\begin{align}
\label{rfake2}
    \widetilde{r}(\bm{x_t}, \bm{a_t}, \bm{x_{t+1}}) := r_{\widetilde{\X}}(\bm{x_t}, \bm{a_t}, \bm{x_{t+1}}) - r_{\widetilde{\X}}(\bm{x_{t-1}}, \bm{a_{t-1}}, \bm{x_{t}}),
\end{align}
where $r_{\widetilde{\X}}$ is the reward in (\ref{rcalib}) but associated to cashfows $\widetilde{\X}$, and $\phi$ the implied volatility mapping associated to $(k_2,t_2)$. One can observe that (\ref{rfake}) is true due to the telescopic property (\ref{rfake2}) and that both $r$ and $\widetilde{r}$ coincide at time $t_2$.

In order to get a more meaningful reward signal, one should subtract to the reward in (\ref{rcalib}) the calibration error obtained with a reference model, for example the Black \& Scholes model or the local volatility model.

We have the following game theoretical interpretation of the calibration problem, which follows by definition of the game reward (\ref{vf2}), (\ref{rcalib}). 
\begin{proposition}
\textbf{(Calibration as a Nash equilibrium)} Let the number of Monte Carlo paths $n$ and time discretization value $\delta$ be fixed. Let $\bm{\pi^*}:=(\pi^*_i)_{i \in [1,n]}$ be a minimizer, over all models $\sigma$ simulated with Monte Carlo parameters $(n,\delta)$, of the calibration error associated to a given set of options. Then $\bm{\pi^*}$ is a Nash equilibrium of the calibration game.
\end{proposition}

Our calibration game is symmetric, so it is common to look for symmetric equilibria, i.e. where all players play the same strategy $\pi^*_i=\pi^*$ $\forall i \in [1,n]$ \cite{vadori:2020:calibration, wellman1,hefti1}. Asymmetric equilibria of symmetric games exist but they represent a degenerate case \cite{FEY2012424}, i.e. they are not natural and one has to work hard to identify them. We discuss in section \ref{sec3} MARL driven methods to find such equilibria.

\textbf{Players's actions $a_t$ and volatility dynamics $\sigma$}. For simplicity, we previously defined the volatility to be players' actions: $a^i_t:=\sigma^i_t$. Note that since actions are sampled from the distribution $\pi_i$, they are by construction stochastic. It is however straightforward to consider a distribution $\varphi \equiv \varphi(\cdot;\bm{z})$ parametrized by (a vector of) parameters $\bm{z}$, and simply define players' actions to be those parameters, i.e. $\sigma^i_t \sim \varphi(\cdot; a^i_t, x^i_t)$. Specific examples of such include the case where $\bm{z}$ are parameters of a Gaussian mixture, possibly correlated with $S$. In that case the overall correlation matrix is of size $n_{mix}+1$ (volatility and spot), where $n_{mix}$ is the number of components in the mixture. The rationale for adding more components is to make volatility increments deviate from Gaussianity, conditional on $x^i_t$. In particular, this includes the case where $a^i_t:=(\mu^i_t,\eta^i_t,\rho^i_t)$ are parameters driving the SDE of $\sigma$, where $Z^{i}_t$ are independent from $Z^{i,\perp}_t$ as in (\ref{lns}):
\begin{align}
\label{actvol}
\ln \sigma^i_{t+1} = \ln \sigma^i_{t} + \mu^i_t \delta + \eta^i_t \rho^i_t \sqrt{\delta} Z^i_t + \eta^i_t \sqrt{1-\rho^{i,2}_t} \sqrt{\delta} Z^{i,\perp}_t.  
\end{align}

It is important to keep in mind that since actions depend on the state $x^i_t$, which as discussed can encapsulate complex information such as path-dependence, $(\mu^i_t,\eta^i_t,\rho^i_t)$ can be complex functions of the state.

\section{Multi-agent reinforcement learning of the calibration game}
\label{sec3}

We discuss in this section a MARL-based method to solve empirically the game presented in section \ref{sec2}. Our main innovation is the concept of basis players needed to handle large number of players $n$, which turned out to be crucial in making our algorithm work. We postpone to future work the connection of our algorithm to mean-field RL based methods. We will consider a policy gradient approach, i.e. where $\pi \equiv \pi_\theta$ is a neural network with parameters $\theta$ that will be iteratively improved.

Our action and state spaces are continuous as subsets of $\R^d$, and we aim at finding symmetric equilibria of the symmetric game, as discussed in section \ref{sec2}, i.e. one where all players play the same strategy. Denoting $v(\pi';\pi)$ the expected utility of player when it plays $\pi'$ while all others play $\pi$, one can find symmetric equilibria by solving:
\begin{align}
    \min_{\pi} \max_{\pi'} v(\pi';\pi) - v(\pi;\pi).
\end{align}
This was done in \cite{wellman1} and takes its origins in \cite{hefti1} under the name SOFA (symmetric opponents form approach). In our context, the $\min$ and $\max$ operations each involve solving a RL problem, so they are costly. The alternative that we will adopt is to approximate $\max_{\pi'}$ by a gradient step. This is a form of self-play, but in the context of $n>2$ players, where $n-1$ players, the "rest of the world", are tied together by a policy $\pi$. This was discussed in \cite{vadori:2020:calibration, vadori:2023:calibration} under the name \textit{shared equilibrium}, where it was shown that policy sharing among players amounts to picking one player at random and making a step towards improving its reward, while freezing other players. In general, it is important to understand that convergence of self-play is linked to transitivity in games and is different, for example, from maximizing the sum of players' rewards. As a simple example, the game where 2 players receive respectively $f(\pi') - f(\pi)$ and $f(\pi) - f(\pi')$ has by construction a sum of players' utilities equal to zero, but optimizing at each training iteration the utility of a randomly selected player will eventually lead to finding a (local) maximizer of $f$.

Recent successes of policy gradient methods are based on two pillars: the policy network $\pi_\theta$, and the value function network $V_\psi$, parametrized by $\psi$. $V_\psi(x^i_t)$ estimates the expected sum of future rewards of an agent currently in state $x_t^i$ and serves as a variance reduction method. The advantage $A_\psi$ represents the benefit of action $a$ in state $x$, relative to some average level in state $x$: 
\begin{align}
\begin{split}
A_\psi(x, a):= &- V_\psi(x)  \\
&+ \EE \left[\sum_{t'=t}^{T} r(\bm{x}_{t'},\bm{a}_{t'},\bm{x}_{t'+1})| x^{i}_{t}=x,a^{i}_{t}=a\right].
\end{split}
\end{align}

The conditional expectation appearing in $A_\psi$ is estimated by a sample of the simulated trajectory, but it could be estimated using truncated trajectories instead (Generalized Advantage Estimation) \cite{gae}. 

The gradient $g_\theta$ according to which the shared policy $\pi_\theta$ is updated is computed by collecting all players' experiences simultaneously and treating them as distinct sequences of local states, actions and rewards experienced by the shared policy \cite{Gupta2017-it,vadori:2020:calibration}, yielding the following expression under A2C (advantage actor critic) \cite{a2c}, where we allow a set of parallel runs $B$:
\begin{align}
\label{grad}
 g_\theta = \underbrace{\frac{1}{n} \sum_{i=1}^{n}}_{\mbox{\scriptsize players i}}
 \underbrace{\frac{1}{B} \sum_{b=1}^{B}}_{\mbox{\scriptsize parallel runs}}\underbrace{\sum_{t=0}^{T} \nabla_{\theta} \ln \pi_{\theta} (a^{i,b}_{t}|x^{i,b}_{t}) \cdot A_\psi(x^{i,b}_{t}, a^{i,b}_{t})}_{\mbox{\scriptsize partial gradient $\partial_i$ of player i's utility}}.
\end{align}
The gradient in equation (\ref{grad}) is presented in the context of A2C, but other policy gradient updates based on $A_\psi$ and $\pi_\theta$ can be used instead, for example we use proximal policy optimization (PPO) in our experiments \cite{ppo}.

The value network $V_\psi$ is estimated classically by performing a few steps of mini-batch gradient descent on the following objective, at each policy update:
\begin{align}
\label{valobj}
\frac{1}{nB} \sum_{i=1}^{n}
\sum_{b=1}^{B} \sum_{t=0}^{T} \left(- V_\psi(x^{i,b}_{t}) + \sum_{t'=t}^{T} r(\bm{x}_{t'}^{b},\bm{a}_{t'}^{b},\bm{x}_{t'+1}^{b})  \right)^2.
\end{align}

\textbf{Basis players.} The gradient update (\ref{grad}) scales with the time horizon $T$ and the number of players $n$. It is clearly not efficient when $n$ is large. In the case of local volatility or particle methods \cite{guyonpa}, one does not compute the leverage function at every timestep of every Monte Carlo path, as it would be highly inefficient. Instead, one computes it on a much smaller grid of size $n_p \sim 50$ in space and interpolate. We take inspiration from this idea to interpolate all players from a smaller set of \textit{basis players}. In practice, a common choice for the action space exploration strategy in policy gradient methods is to let $\pi_{\theta}=(\pi^\mu_{\theta},\pi^\sigma_{\theta})$ output the mean and standard deviation of actions:
\begin{align}
\label{bpl}
\begin{split}
& a^i_t = \pi_\theta(x^i_t) = \pi^\mu_\theta(x^i_t) +  Z^{\pi}_{i,t} \cdot \pi^\sigma_\theta(x^i_t),\\
& Z^{\pi}_{i,t} \sim N(0,1), \hspace{2mm} i \in [1, n_p].
\end{split}
\end{align}

$\pi^\mu_{\theta}$ and $\pi^\sigma_{\theta}$ are natural interpolators, so $\pi^\mu_{\theta}(x^j_t)$ and $\pi^\sigma_{\theta}(x^j_t)$ can simply be computed by the forward pass of a neural network and as such can be evaluated cheaply for all trajectories $j$. The \textit{exploration} variable $Z^{\pi}_{i,t}$ represents how players explore the space of actions and find "good" and "bad" actions, by using (\ref{grad}) to update their knowledge. Without $Z^{\pi}$, i.e. $Z^{\pi} \equiv 0$, equation (\ref{grad}) would not lead to any policy improvement. It is the analogous of the exploration policy used in $Q-$learning, typically $\epsilon$-greedy. Our innovation is to compute action $a^j_t$ of each trajectory $j$ as:
\begin{align}
\label{bpl2}
\begin{split}
&a^j_t :=\pi^\mu_\theta(x^j_t) + \widehat{Z}^{\pi}_{j,t} \pi^\sigma_\theta(x^j_t),\\
&\widehat{Z}^{\pi}_{j,t}=\mbox{interp}[x^i_t, Z^{\pi}_{i,t}]_{i \in [1, n_p]}, \hspace{2mm} j \in [1, n],
\end{split}
\end{align}
where $\widehat{Z}^{\pi}_{j,t}$ is interpolated from the $n_p$ basis players' exploration $(Z^{\pi}_{i,t})_{i \in [1,n_p]}$. This way, exploration performed by basis players is propagated to the whole Monte Carlo computation, as each basis player impacts other players in its neighborhood via interpolation. As an interpolation method, we tried linear interpolation and $k$-nearest neighbors ($k$-NN), and both yielded similar results \footnote{LinearNDInterpolator and KNeighborsRegressor in SciPy.}. Basis players are picked at random among the $n$ trajectories at every time $t$. We summarize our approach in algorithm \ref{algov}.

\begin{algorithm}[ht]
\caption{\textbf{MARLVol}}\label{algov}
\textbf{Input:} {learning rate $\alpha$, initial policy and value  network parameters $\theta$, $\psi$, number of basis players $n_p$.}
\begin{algorithmic}[1]
\While{game reward not converged}
    \For{each parallel run $b \in [1,B]$, time $t \in [0,T]$}
    \State{pick $n_p$ basis players at random among $[1,n]$.}
    \State{sample actions $a^{i,b}_{t} \sim \pi_{\theta}$ for basis players $i \in [1,n_p]$ as in (\ref{bpl}).}
    \State{interpolate actions $a^{j,b}_{t}$ for all trajectories $j \in [1,n]$ as in (\ref{bpl2}).}
    \State{compute $\sigma^{j,b}_t \sim \varphi(\cdot; a^{j,b}_t, x^{j,b}_t)$ for all trajectories $j \in [1,n]$ as discussed in section \ref{sec2}.}
    \State{diffuse $S^{j,b}_{t}$ for all trajectories $j \in [1,n]$ as in (\ref{lns}).}
    \State{compute reward $r^{j,b}_{t}$ for all trajectories $j \in [1,n]$ as in (\ref{rcalib}).}
    \EndFor
    \State{compute $g_\theta$ as in (\ref{grad}) with trajectories from the $n_p$ basis players.}
     \State{update policy network $\theta = \theta + \alpha g_\theta$ and update value network $V_\psi$ by performing a few steps of gradient descent on the objective (\ref{valobj}).}
\EndWhile
\Return $\theta$
\end{algorithmic}
\end{algorithm}

\section{Experiments}
\label{secexp}

We consider S\&P 500 (SPX) smiles as of August 1st 2018 for $t_1=21$ days and $t_2=51$ days. Smiles for dates in between are interpolated. 

\begin{figure}[ht]
    \centering
    \begin{subfigure}[b]{0.32\textwidth}
        \centering
        \includegraphics[width=\textwidth]{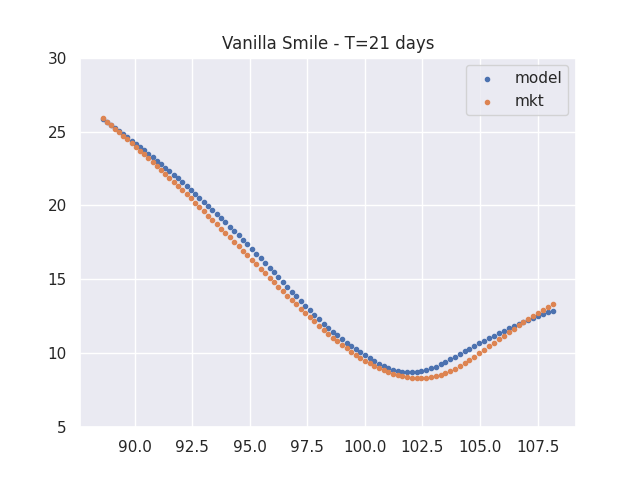}
    \end{subfigure}
    \hfill
    \begin{subfigure}[b]{0.32\textwidth}
        \centering
        \includegraphics[width=\textwidth]{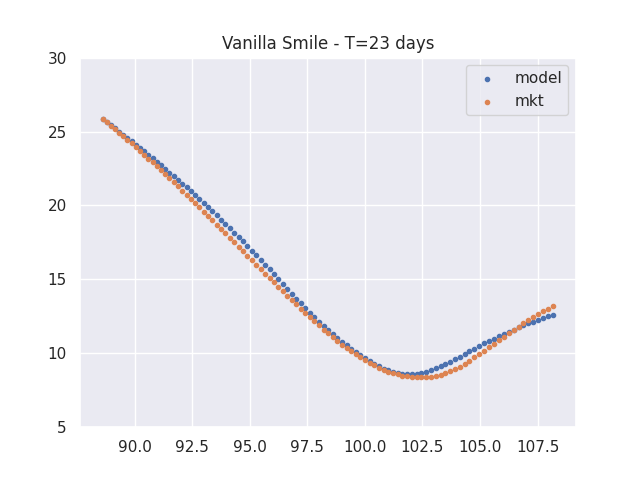}
    \end{subfigure}
    \begin{subfigure}[b]{0.32\textwidth}
        \centering
        \includegraphics[width=\textwidth]{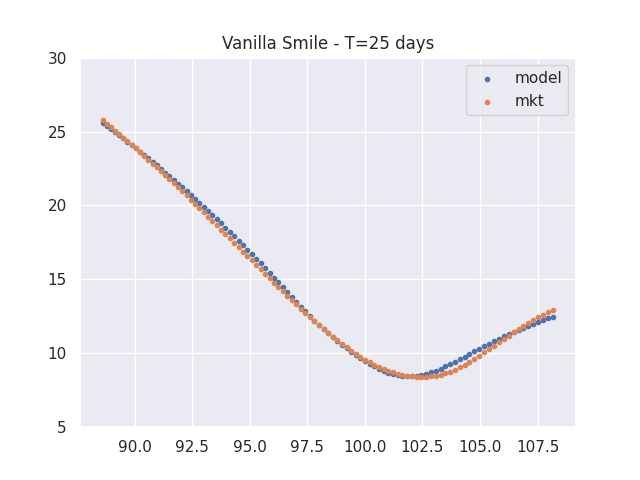}
    \end{subfigure}
    \begin{subfigure}[b]{0.32\textwidth}
        \centering
        \includegraphics[width=\textwidth]{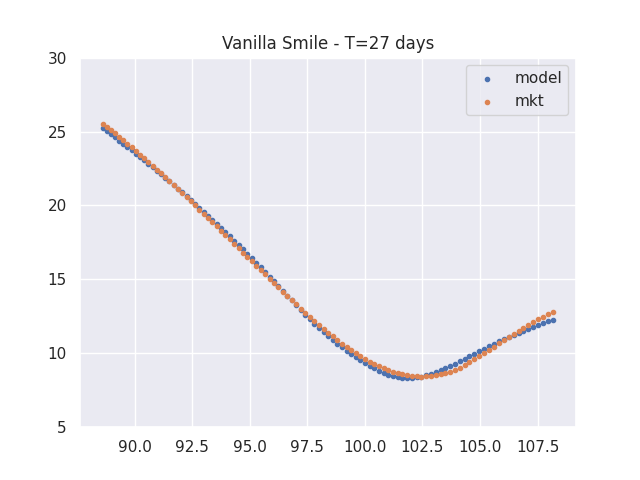}
    \end{subfigure}
    \hfill
    \begin{subfigure}[b]{0.32\textwidth}
        \centering
        \includegraphics[width=\textwidth]{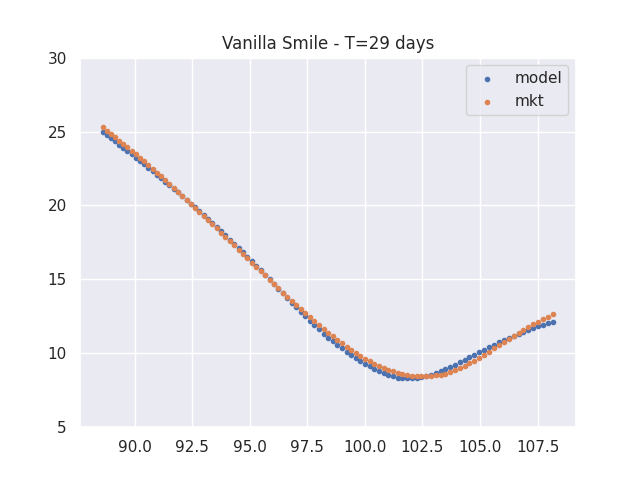}
    \end{subfigure}
    \begin{subfigure}[b]{0.32\textwidth}
        \centering
        \includegraphics[width=\textwidth]{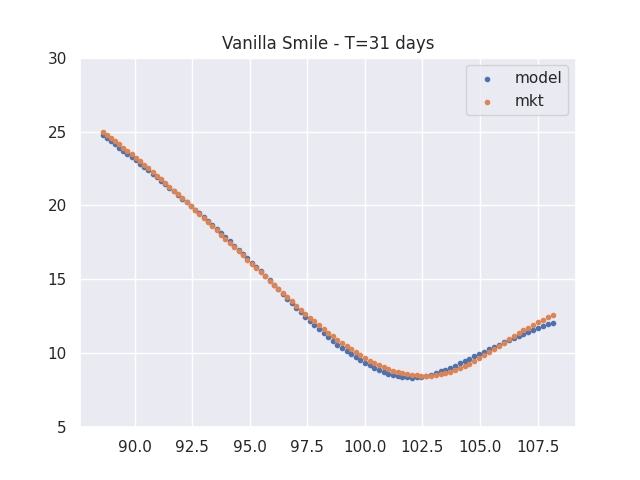}
    \end{subfigure}
    \begin{subfigure}[b]{0.32\textwidth}
        \centering
        \includegraphics[width=\textwidth]{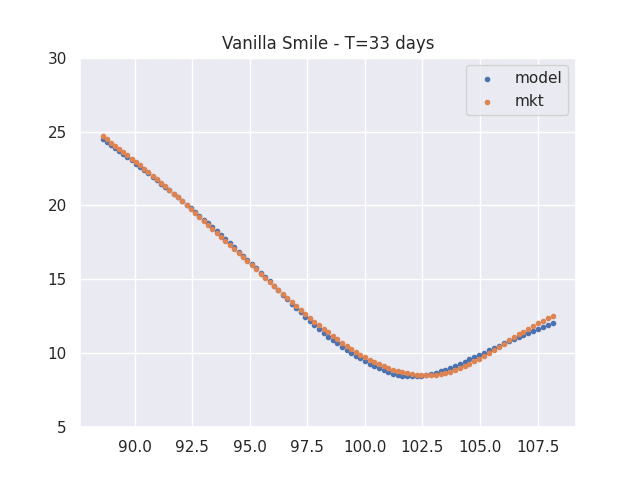}
    \end{subfigure}
    \hfill
    \begin{subfigure}[b]{0.32\textwidth}
        \centering
        \includegraphics[width=\textwidth]{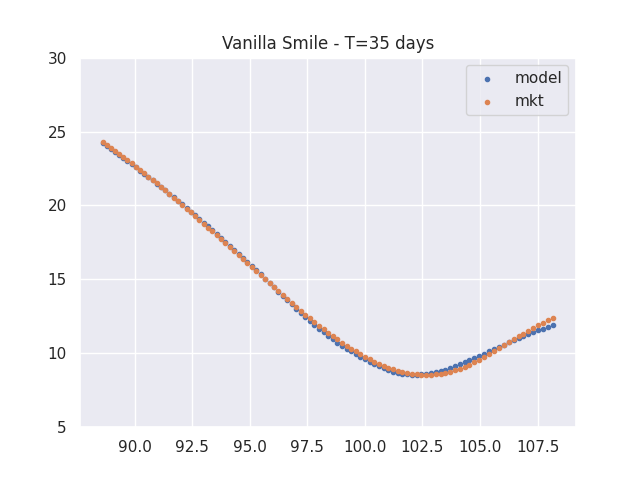}
    \end{subfigure}
    \begin{subfigure}[b]{0.32\textwidth}
        \centering
        \includegraphics[width=\textwidth]{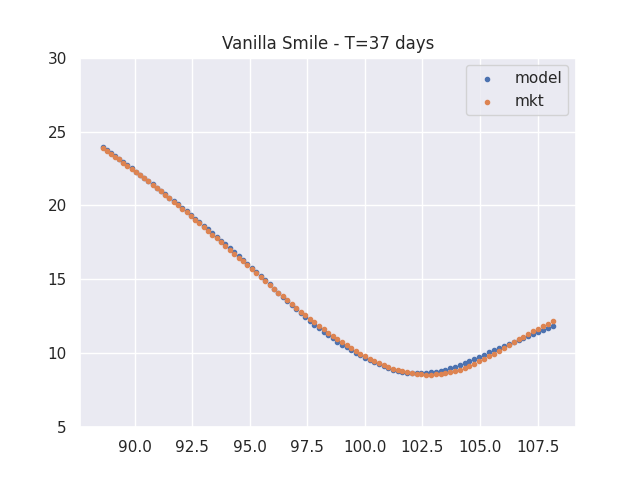}
    \end{subfigure}
    \begin{subfigure}[b]{0.32\textwidth}
        \centering
        \includegraphics[width=\textwidth]{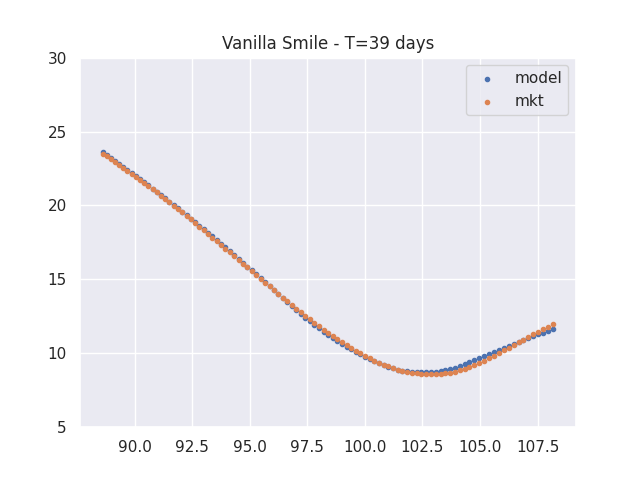}
    \end{subfigure}
    \hfill
    \begin{subfigure}[b]{0.32\textwidth}
        \centering
        \includegraphics[width=\textwidth]{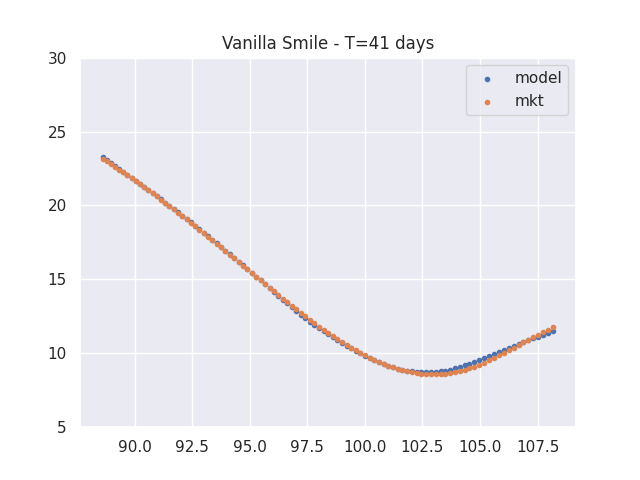}
    \end{subfigure}
    \begin{subfigure}[b]{0.32\textwidth}
        \centering
        \includegraphics[width=\textwidth]{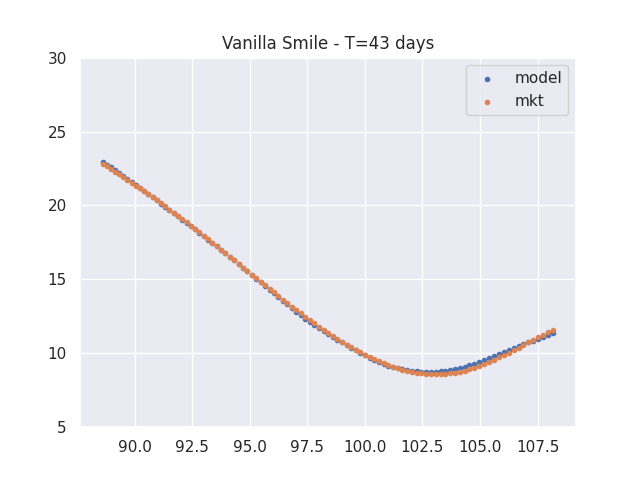}
    \end{subfigure}
    \begin{subfigure}[b]{0.32\textwidth}
        \centering
        \includegraphics[width=\textwidth]{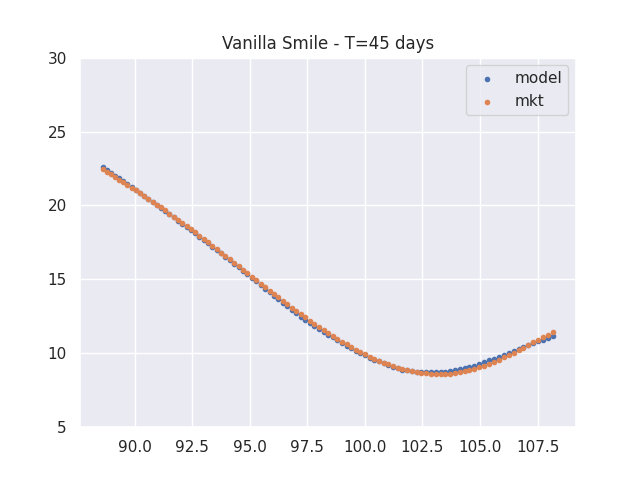}
    \end{subfigure}
    \hfill
    \begin{subfigure}[b]{0.32\textwidth}
        \centering
        \includegraphics[width=\textwidth]{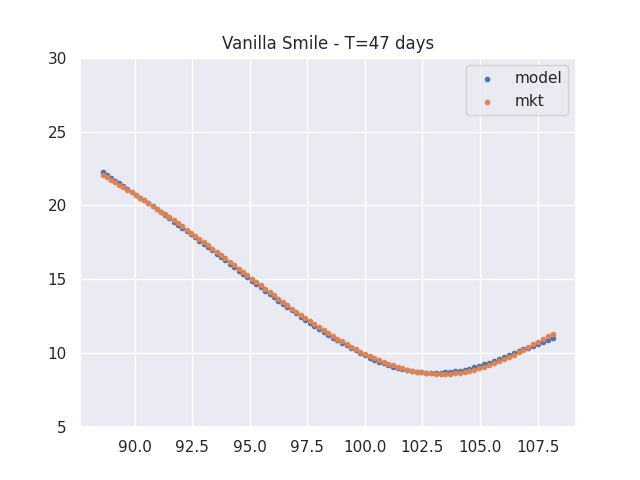}
    \end{subfigure}
    \begin{subfigure}[b]{0.32\textwidth}
        \centering
        \includegraphics[width=\textwidth]{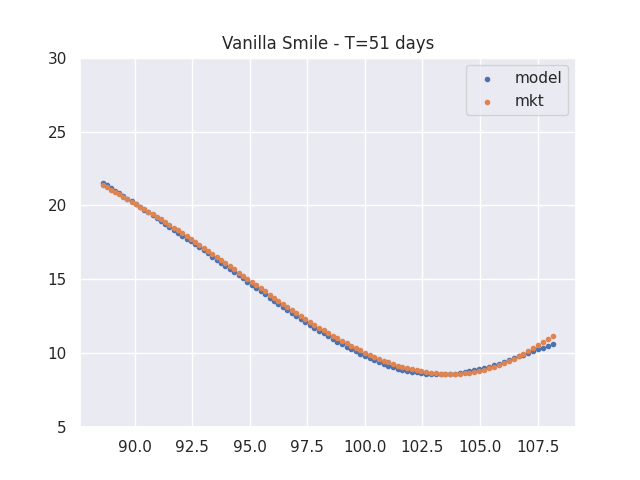}
    \end{subfigure}
    \caption{continuum of learnt vanilla smiles from $t_1=21$ days to $t_2=51$ days. Y-axis: option price in implied volatility points, X-axis: strike as percentage of the spot. SPX, August 1st 2018.}
    \label{figvanilla}
\end{figure}

We first learn local volatility: our target instruments are all call options for all maturities up to $t_2$. We see in figure \ref{figvanilla} that we are able to fit the continuum of market smiles. In this case $x_t^i$ consists of $(t,S^i_t)$, and actions $a^i_t$ consist of the mean and variance of a normal distribution, which we sample $\ln \sigma_t^i$ from. We trained using a time step $\delta=1$ day ($\delta = \frac{1}{252}$), so that $T=51$, $n=1.2 \cdot 10^5$ trajectories, $B=10$ parallel runs and $n_p=100$ basis players. We checked empirically that the results are not too sensitive to these parameters, for example we considered $T=102$ ($\delta=0.5$ day), $n=2.4 \cdot 10^5$, and $n_p$ from 20 to 100. Our approach doesn't suffer too much from increasing these parameters except from the incompressible cost of the Monte Carlo simulation, since we do not need to store the computational graph as discussed in section \ref{sec1}.

\begin{figure}[ht]
    \centering
    \begin{subfigure}[b]{0.34\textwidth}
        \centering
        \includegraphics[width=\textwidth]{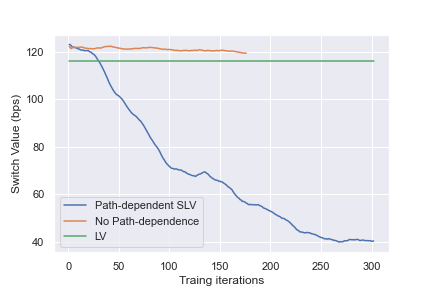}
    \end{subfigure}
    \begin{subfigure}[b]{0.32\textwidth}
        \centering
        \includegraphics[width=\textwidth]{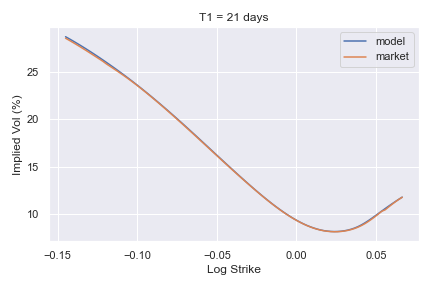}
    \end{subfigure}
    \hfill
    \begin{subfigure}[b]{0.32\textwidth}
        \centering
        \includegraphics[width=\textwidth]{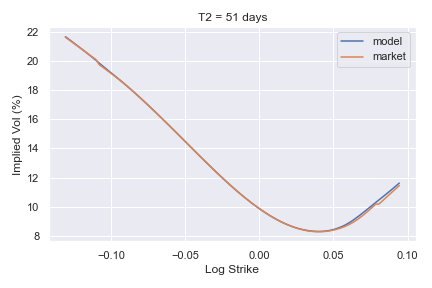}
    \end{subfigure}
    \caption{(Left) Bermudan option price with exercise dates $t_1=21$ days, $t_2=51$ days during training. Path dependence of the volatility process via the knowledge of $(S_{t_1},\sigma_{t_1})$ is required to minimize the price. (Middle and right) we sanity check the vanilla smiles obtained via localization of the learnt path-dependent volatility (Gy\"{o}ngy). SPX, August 1st 2018.}
    \label{figb1}
\end{figure}

\begin{figure}[ht]
    \centering
    \begin{subfigure}[b]{0.49\textwidth}
        \centering
        \includegraphics[width=\textwidth]{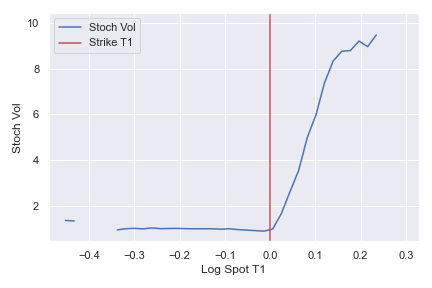}
    \end{subfigure}
    \hfill
    \begin{subfigure}[b]{0.49\textwidth}
        \centering
        \includegraphics[width=\textwidth]{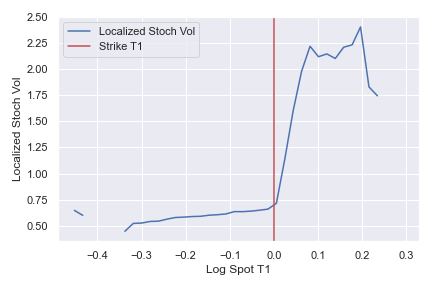}
    \end{subfigure}
    \caption{Bermudan option with exercise dates $t_1=21$ days, $t_2=51$ days. (Left) average volatility $\sigma_t$ and (right) average localized volatility $\EE[\sigma_t^2|S_t]^{-\frac{1}{2}} \sigma_t$ as a function of $\ln S_{t_1}$. Strike $\ln k_1=0$. We learn to allocate higher volatility on the region $S_{t_1}>k_1$ where the $t_1$ call option has non zero payoff. Learnt volatility is an increasing function of $S_{t_1}$, matching our intuition. SPX, August 1st 2018.}
    \label{figb3}
\end{figure}

We then turn to the Bermudan$(t_1,t_2)$ case for which we have $\X:= \max \{(S_{t_1} - k_1)^+,\EE_{t_1}[(S_{t_2} - k_2)^+]\}$. We choose strikes $k_1=S_0=1$, $k_2 = 1.012$, so that the vanilla calls $(t_1,k_1)$, $(t_2,k_2)$ have the same value: this is the interesting case, since if one call is worth much more than the other, the Bermudan price will already be close to its maximum european lower bound. We define the switch value as the difference between the Bermudan and the maximum european prices, both expressed in implied volatility points associated to $(t_2,k_2)$. By construction the switch value is nonnegative. We consider similar numerical parameters as in the local volatility case, and considered different ways to define actions as discussed in section \ref{sec2} and in (\ref{actvol}), but our conclusions remained similar. The option price is calculated using American Monte Carlo \cite{ls} with polynomial regression of degree 8. We compare two cases regarding the construction of $x_t^i$ (i) the path-dependent case considers $x_t^i=(t, S^i_t, \sigma^i_{t-1}, S^i_{t \wedge t_1}, \sigma^i_{(t-1) \wedge t_1})$, and (ii) the non path-dependent case considers $x_t^i = (t, S^i_t, \sigma^i_{t-1})$. We see in figure \ref{figb1} that path-dependence is required to minimize the price, while remaining calibrated to vanillas. Note that in this case, our calibration reward only consists of the Bermudan objective, while we ensure calibration to vanillas for all dates $t \in [t_1,t_2]$ using localization. We check in figures \ref{figb2}-\ref{figb3} the nature of the learnt path-dependence, and we see that as intuited in section \ref{sec1}, we learn a volatility increasing in $S_{t_1}$, decreasing in $S_{t}|_{t>t_1}$. In figure \ref{figb3}, we see that the cutoff strike $k_1$ is perfectly learnt by our RL policy, as it is obviously never optimal to exercise at $t_1$ if $S_{t_1}<k_1$.

Due to the flexibility in defining the state $x^i_t$, our approach can be extended straightforwardly to more Bermudan dates, where path-dependent information corresponding to past exercise dates is fed into the state process.

The RL policy $\pi_\theta$ was trained in the RLlib framework \cite{rllib}, and run on AWS using a EC2 C5 4xlarge instance with $B=10$ CPUs, using Proximal Policy Optimization \cite{ppo}. We used configuration parameters in line with the latter paper, that is a clip parameter of 0.3, an adaptive KL penalty with a KL target of $0.01$ and a learning rate of $10^{-4}$. Each policy update was performed with 30 iterations of stochastic gradient descent with SGD minibatch size of $0.1 \cdot n_p \cdot T \cdot B$. We used a fully connected neural net with 3 hidden layers, 50 nodes per layer, and $\tanh$ activation. Although $\tanh$ is much more expensive to evaluate than ReLU, we noticed that it lead to smoother and more efficient learning.

\section{Conclusion}
\label{secc}
In this paper we provided a novel view on the derivative pricing model calibration problem using game theory and multi-agent reinforcement learning. We illustrated its ability to learn local volatility, as well as path-dependence in the volatility process required to minimize Bermudan option prices. 

\clearpage

\clearpage

\section*{Disclaimer}
This paper was prepared for information purposes by the Artificial Intelligence Research group of JPMorgan Chase \& Co and its affiliates (“JP Morgan”), and is not a product of the Research Department of JP Morgan. JP Morgan makes no representation and warranty whatsoever and disclaims all liability, for the completeness, accuracy or reliability of the information contained herein. This document is not intended as investment research or investment advice, or a recommendation, offer or solicitation for the purchase or sale of any security, financial instrument, financial product or service, or to be used in any way for evaluating the merits of participating in any transaction, and shall not constitute a solicitation under any jurisdiction or to any person, if such solicitation under such jurisdiction or to such person would be unlawful.

\small
\bibliography{n20}
\bibliographystyle{apalike}
\normalsize

\setcounter{section}{0}
\appendix
\renewcommand{\thesection}{\Alph{section}}


\end{document}